\documentclass[reprint,aps,prl,superscriptaddress,amsmath,amssymb,floatfix,footinbib,longbibliography]{revtex4-1}
\usepackage{graphicx}
\usepackage{dcolumn}
\usepackage{bm}
\usepackage{amsmath}
\usepackage{times}
\usepackage{color}
\usepackage[breaklinks=true,colorlinks,citecolor=blue,linkcolor=blue,urlcolor=blue]{hyperref}
\usepackage{ragged2e}

\makeatletter

\newcommand{\Rmnum}[1]{\expandafter\@slowromancap\romannumeral #1@}
\makeatother

\begin{document}

\title{Spin-reorientation driven topological Hall effect in Fe$_4$GeTe$_2$}
\author{Alapan Bera}
\affiliation{Department of Physics, Indian Institute of Technology Kanpur, Kanpur 208016, India}
\author{Soumik Mukhopadhyay}
\email{soumikm@iitk.ac.in}
\affiliation{Department of Physics, Indian Institute of Technology Kanpur, Kanpur 208016, India}

\begin{abstract}
Iron-based van der Waals (vdW) ferromagnets with relatively high ordering temperatures are a current research focus due to their significance in fundamental physics and potential applications in spintronics. Competing magnetic interactions and anisotropies can give rise to nontrivial spin textures in these materials, resulting in novel topological features. Fe$_4$GeTe$_2$ (F4GT) is a nearly room-temperature vdW ferromagnet, well known for hosting a spin-reorientation transition (SRT) arising from the interplay of perpendicular magnetic anisotropy (PMA) and shape anisotropy. In this work, we investigate the angle-dependent magneto-transport properties of F4GT single crystals. We report a large topological Hall effect (THE) in a multi-layer F4GT originating from the SRT-driven non-coplanar spin textures. The THE appears at the in-plane orientation of the external magnetic field and persists over a wide range of temperatures around SRT. Additionally, we find a thickness-sensitive THE signal for the c axis orientation of the magnetic field at a low-temperature regime which is associated with a reentrant Lifshitz transition.
\end{abstract}
\maketitle

{\it Introduction.---} 
In the last few decades, intensive research focusing on the topological aspect of condensed matter physics has led to many remarkable and exciting discoveries. Novel topological materials, such as Weyl and Dirac semimetals, topological insulators, and topological superconductors, have emerged as platforms for exploring unique and exotic physical phenomena. Since the discovery of the skyrmion lattice in the chiral ferromagnet MnSi~\cite{Sk_MnSi}, observation of numerous other topologically non-trivial chiral spin textures, such as antiskyrmions~\cite{AntiSk}, skyrmionium~\cite{Skyrmionium}, (anti)merons~\cite{Meron_FGT, AntiMeron}, bimerons~\cite{Bimeron_1}, etc. have been reported in various other systems. A finite scalar spin chirality typically characterizes these topological vortices and other non-coplanar spin textures. As electrons hop between localized spin sites with scalar spin chirality, they acquire a Berry phase factor, resulting in a fictitious magnetic field. The emergent gauge field leads to intriguing magneto-transport phenomena, such as the topological Hall effect (THE). The Berry phase factor is also responsible for the well-known anomalous Hall effect (AHE) in ferromagnets. However, unlike the AHE, which arises from an electron acquiring an anomalous velocity due to a finite Berry phase in \textbf{k}-space, the THE originates from Berry phase accumulation in real space caused by chiral spin textures. THE originating from topological spin textures has mostly been observed in non-centrosymmetric systems like MnSi~\cite{THE_MnSi}, MnGe~\cite{THE_MnGe}, Mn$_5$Si$_3$~\cite{THE_Mn5Si3}, FeGe~\cite{THE_FeGe}, etc, as a broken inversion symmetry favors Dzyaloshinsky-Moriya interaction (DMI). There also exists experimental observation of THE in a few centrosymmetric systems such as MnNiGa~\cite{THE_MnNiGa}, NdMn$_2$Ge$_2$~\cite{THE_NdMn2Ge2}, Mn$_3$Sn~\cite{THE_Mn3Sn}, Gd$_2$PdSi$_3$~\cite{THE_Gd2PdSi3}, etc., attributed to magnetic frustration induced local breaking of inversion symmetry. 

Van der Waals (vdW) magnets have garnered significant attention in the field of low-dimensional magnetism in recent years, owing to their atomically smooth surface which can be conveniently integrated into novel heterostructures. Fe$_\mathrm{n}$GeTe$_2$ (n= 3,4,5) is a family of vdW ferromagnets that is well-known for their high Curie temperature, which extends up to 310 K in Fe$_5$GeTe$_2$~\cite{May2019}. Competing magnetic interaction and anisotropies often lead to exotic magnetic phases and topologically non-trivial spin formations in these systems. Recent observations of skyrmions and merons in centrosymmetric Fe$_3$GeTe$_2$~\cite{Sk_F3GT, Sk_F3GT2}, Fe$_5$GeTe$_2$~\cite{Schmitt2022}, Fe$_{\mathrm{5-x}}$GeTe$_2$~\cite{Sk_F5GT, Meron_FGT} and Fe$_3$GaTe$_2$~\cite{FGaT1, FGaT2, FGaT3, FGaT4} holds promise for next-generation energy-efficient and high-performance technologies. On the other hand, Fe$_4$GeTe$_2$ (F4GT) is a near room-temperature vdW ferromagnet of the same family with a well-known spin reorientation transition (SRT) at the intermediate temperature which arises from the competing magnetic anisotropies~\cite{F4GT_1, F4GT_2}. Despite belonging to the centrosymmetric $R\overline{3}m$ space group, these competing interactions can potentially give rise to distinctive topological properties.

In this work, we study the angle-dependent magneto-transport properties of F4GT crystal to characterize its topological features. We identify a large THE near the SRT temperature for the ab plane orientation of the applied magnetic field. The THE signature persists down to the lowest measured temperature of 2 K with diminishing strength. For c axis orientation of the DC field, we observe a thickness-dependent THE signature at low temperatures associated with a reentrant Lifshitz transition.  

{\it Experimental methods.---}
F4GT single crystals are grown using the chemical vapor transport (CVT) method with iodine (I$_2$) as a transport agent. The typical crystal dimension is approximately 2 $\times$ 2 $\times$ 0.1 mm$^3$. Energy dispersive X-ray spectroscopy (EDS) measurement is carried out on freshly cleaved crystal surfaces to confirm the atomic composition of the F4GT crystals. X-ray diffraction (XRD) measurement is performed using a PANalytical X’Pert diffractometer to study the crystallinity of the bulk single crystals. The static magnetic properties of the bulk samples are studied with a vibrating sample magnetometer (VSM) from Quantum Design Ltd. For the magneto-transport measurements, we use a variable temperature insert (VTI) along with a 12 T superconducting magnet provided by Cryogenics Ltd.  

\begin{figure}
\includegraphics[width=\linewidth]{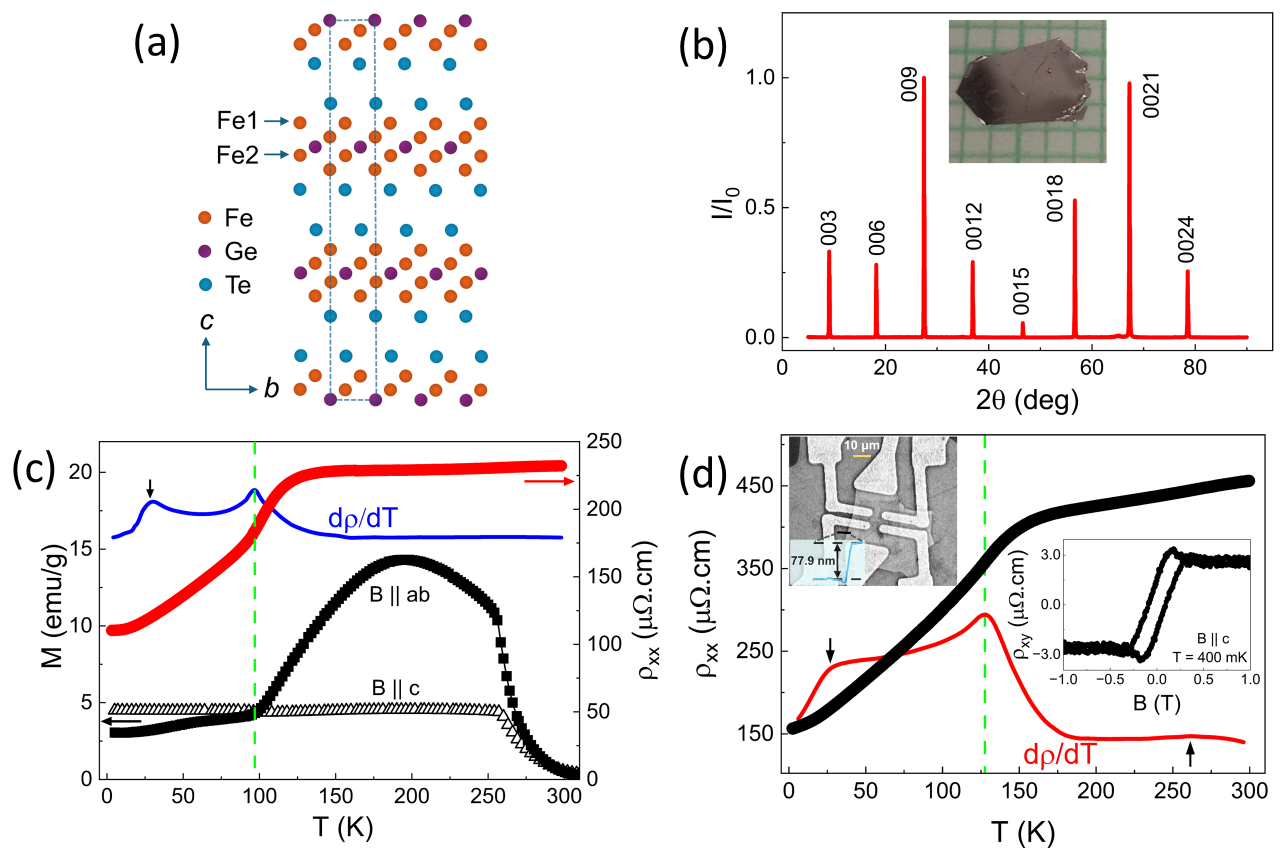}
\caption{(a) Crystal structure of F4GT (side view), with the unit cell outlined by a dashed box. (b) Normalized (00l) peaks obtained from single crystal XRD show good crystallinity of the bulk F4GT samples. The inset shows the optical image of a representative bulk single crystal. (c) Temperature dependence of field-cooled DC magnetization of bulk F4GT at B = 0.05 T (symbols in black). Symbols in red show the temperature dependence of electrical resistivity of bulk F4GT with its temperature derivative being represented by the line in blue. (d) The temperature dependence of the electrical resistivity of multi-layer F4GT (device D1), with its temperature derivative shown in the red line. Inset: SEM image of device D1 (left), Hall resistivity curve at 400 mK (right). The green dashed line marks the prominent SRT transition.}
\label{fig1}
\end{figure} 

{\it Results and discussion.---}
F4GT crystallizes into centrosymmetric R$\bar{3}$m space group with rhombohedral lattice symmetry where Fe$_4$Ge slabs remain sandwiched between Te layers [Fig.~\ref{fig1}(a)]. The XRD pattern of a bulk single crystal in Fig.~\ref{fig1}(b) shows the (00l) peaks confirming the high crystallinity of the sample with the c axis being perpendicular to the crystal ab plane. The inset of Fig.~\ref{fig1}(b) shows the optical image of a typical bulk single crystal. Additionally, the EDS spectra (see Sec. I of the SM~\cite{SM} for the details) obtained from a freshly cleaved bulk crystal exhibit prominent Fe, Ge, and Te peaks, with a stoichiometric ratio close to 4:1:2.  

The temperature-dependent field-cooled DC magnetization ($M-T$) of a bulk F4GT single crystal is shown in Fig.~\ref{fig1}(c) for an external magnetic field of 0.05 T applied along the crystal ab plane and c axis (presented by closed and open symbols, respectively). The significant decrease in magnetization as the temperature approaches 260 K from below suggests a transition from ferromagnetic to paramagnetic behavior. The crossover between ab plane and c axis magnetization near 100 K marks the well-known SRT arising from the competing temperature-dependent anisotropies (magneto-crystalline and shape anisotropy). Effectively, F4GT evolves from an easy-axis (c axis) ferromagnet below 100 K to an easy-plane (ab plane) ferromagnet above 100 K. The temperature-dependence of electrical resistivity ($R-T$) of a bulk crystal is studied using the standard four-probe method. Fig.~\ref{fig1}(c) shows the metallic nature of the resistivity curve with a clear imprint of the SRT near 100 K. Apart from this, we observe another transition [marked by the blue arrow in Fig.~\ref{fig1}(c)] with a hump-like feature in the temperature derivative of the resistivity data, the origin of which is, prima facie, not entirely understood. However, this plays an important role in the magneto-transport properties of F4GT, which will be discussed in more detail later on.

\begin{figure}
\includegraphics[width=\linewidth]{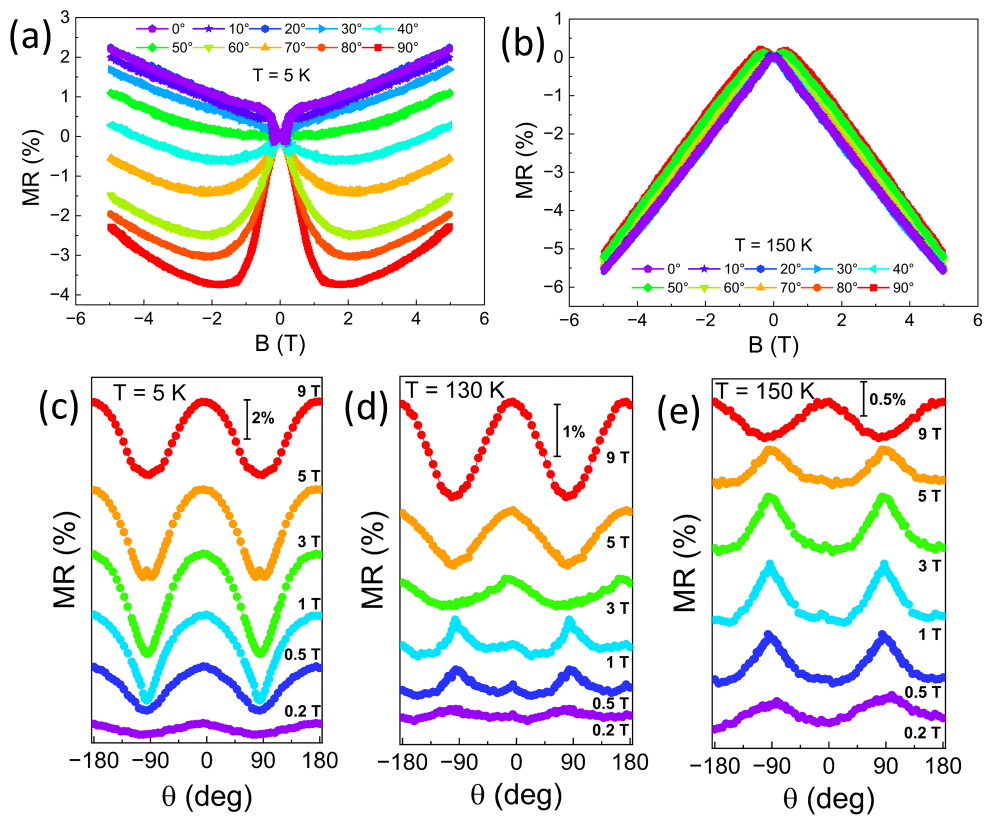}
\caption{MR vs. B curves at (a) 5 K and (b) 150 K for different out-of-plane orientations of the external magnetic field. Anisotropic MR at (c) 5 K, (d) 130 K and (e) 150 K is presented for various magnetic field strengths. At low magnetic field strength (below 9 T), maximum and minimum positions in the MR curve exhibit a shift due to the change in the anisotropy direction caused by SRT as the temperature increases from 5 K to 150 K. The easy axis is along the c axis at 5 K and shifts to the ab plane at 150 K.}
\label{fig2}
\end{figure}

To study the electrical transport properties of a multi-layer F4GT, we fabricate a Hall bar device using e-beam lithography (EBL). A mechanically exfoliated F4GT flake with a uniform thickness of 78 nm is transferred onto pre-patterned electrodes (device D1) on a SiO$_2$/Si substrate (see Sec. III of the SM~\cite{SM} for the details). Scanning electron microscopy (SEM) image of the device is shown in the inset Fig.~\ref{fig1}(d). Fig.~\ref{fig1}(d) shows the temperature-dependent electrical resistivity (see symbols in black) of the multi-layer system. The temperature derivative of resistivity [see the line in red in Fig.~\ref{fig1}(d)] shows the SRT temperature ($\mathrm{T_{SRT}}$) being elevated to $\sim$ 127 K when compared to the bulk counterpart, which is a result of increased perpendicular magnetic anisotropy (PMA) strength at the lower thickness and is consistent with the earlier reports~\cite{F4GT_1, F4GT_2}. However, the low-temperature transition persists at around 30 K (T$^*$), indicating a robust and intrinsic origin. 

\begin{figure*}
\includegraphics[width=0.75\linewidth]{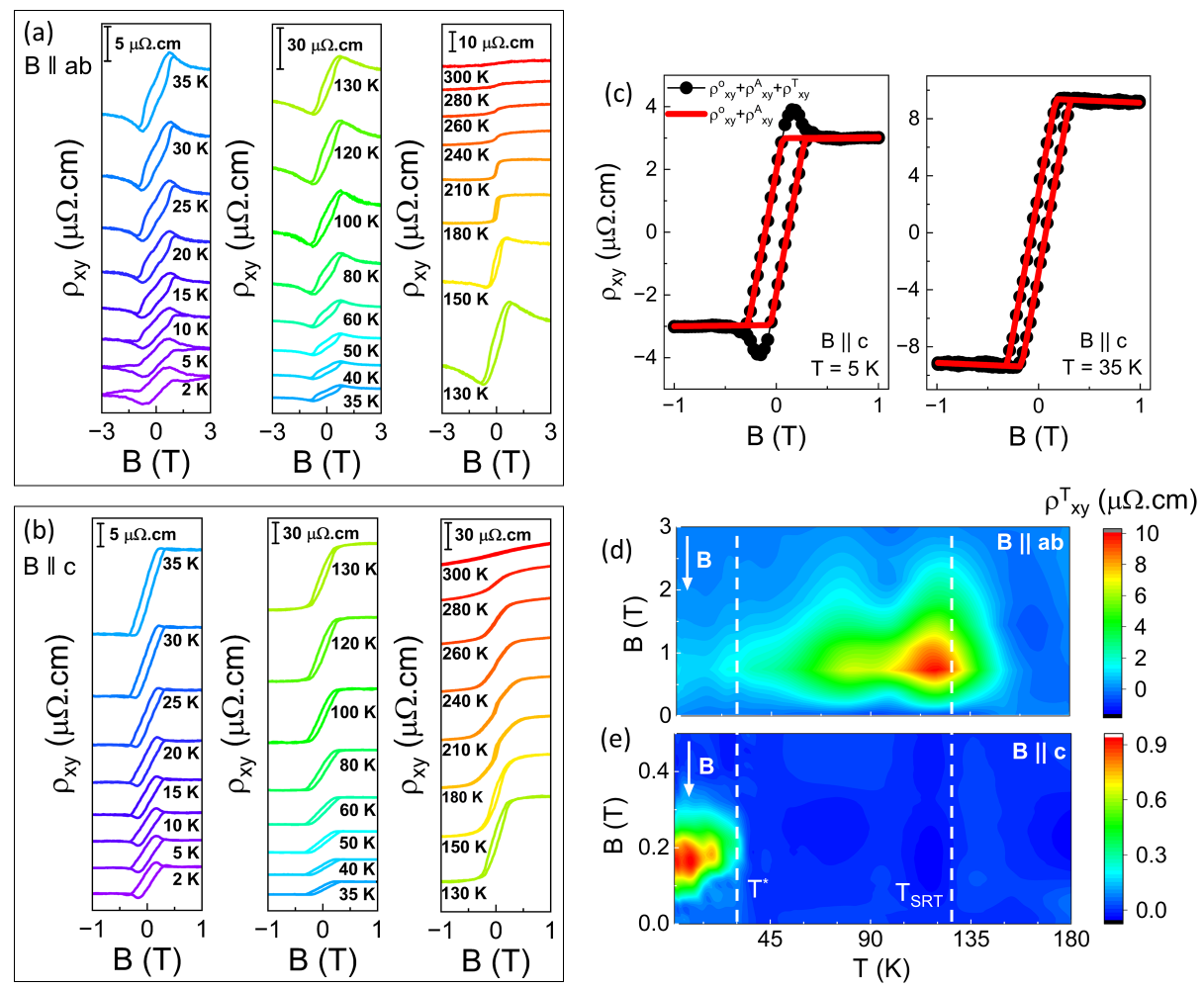}
\caption{\justifying{The field dependence of Hall resistivity with increasing temperature are presented for (a) B$\parallel$ab and (b) B$\parallel$c in F4GT. The curves are stacked along the y-axis for better visualization. (c) Hall curves for B$\parallel$c at 5 K (left) and 35 K (right) with a step function model fit accounting for AHE and OHE contribution. A cusp-like signature is visible at 5 K due to the topological Hall contribution. Contour plots of the THE contribution are presented for (d) B$\parallel$ab and (e) B$\parallel$c as a function of magnetic field and temperature. The white dashed lines mark the temperatures corresponding to the unknown transition and SRT.}}
\label{fig3}
\end{figure*} 

Next, to characterize the magneto-transport properties of the multi-layer F4GT, we measure the field dependence of longitudinal magneto-resistivity (MR) $\rho_{\mathrm{xx}}$ and transverse Hall resistivity $\rho_{\mathrm{xy}}$ simultaneously within a wide temperature range of 400 mK to 300 K at different orientations of the externally applied magnetic field $\mathrm{B}$. The raw MR and Hall resistivity curves were symmetrized and antisymmetrized, respectively, to eliminate the effects of electrode misalignment. Inset of Fig.~\ref{fig1}(d) shows a typical field-dependent $\rho_{\mathrm{xy}}$ curve at 400 mK for B$\parallel$c. The nonlinear behavior of $\rho_{\mathrm{xy}}$, with strong hysteresis, clearly indicates the anomalous Hall effect (AHE) contributing predominantly to the Hall resistivity, the contribution from the ordinary Hall effect (OHE) being negligible due to the metallic nature of F4GT.

Fig.~\ref{fig2}(a) and ~\ref{fig2}(b) presents evolving $\rho_{\mathrm{xx}}$ vs. B curves with polar angle $\theta$ variation at T = 5 K and at T = 150 K, respectively. We define $\theta$ as the angle between the external magnetic field direction and the c axis (normal to the crystal ab plane). The percentage change in magneto-resistance is calculated as MR\% = $\mathrm{\frac{\rho_{xx}(B)-\rho_{xx}(0)}{\rho_{xx}(0)}\times 100\%}$. The high field positive slope of the MR curve is observable for all the polar angles at 5 K. Notably, the MR at a particular field increases as we go from B$\parallel$ab ($\theta=90^\circ$) to B$\parallel$c ($\theta=0^\circ$) configuration which is the opposite behavior of that at 150 K. This is more clearly visible in the anisotropic MR (AMR) measured at temperatures T = 5 K, 130 K, and 150 K as presented in Fig.~\ref{fig2}(c)-(e), respectively. The maximum and minimum positions in the MR curve assume a $90^\circ$ shift as we go from 5 K to 150 K. This provides direct evidence of the shift in anisotropy direction, leading to the SRT, with the c axis as the easy axis at 5 K and the ab plane as the easy plane at 150 K. Refer to Sec. IV of the SM~\cite{SM} for the details on temperature evolution of MR curves.

Next, we present the field dependence of Hall resistivity for B$\parallel$ab and B$\parallel$c in Fig.~\ref{fig3}(a) and ~\ref{fig3}(b), respectively. A prominent anomalous Hall signal is observed up to 240 K in the c axis Hall data, suggesting that the $\mathrm{T_c}$ of the thin flake lies in the range of 240–260 K. A comparatively large saturation field is observed in the ab plane Hall data, with the c axis being the easy axis at temperatures below 130 K. When the magnetic field is oriented along the ab plane, as we go below 150 K, a cusp-like anomaly arises in the low-field region of the $\rho_{\mathrm{xy}}$ vs B curves, indicating the presence of a THE, which persists in the Hall data at temperatures down to 2 K. The THE amplitude increases as the temperature goes up from 2 K and becomes the most prominent in the vicinity of the SRT. On the other hand, For B$\parallel$c no such appearance of the THE is recorded around the SRT and even down to 30 K. Below 30 K, a hump-like THE signal appears in the $\rho_{\mathrm{xy}}$ curve [Fig.~\ref{fig3}(b)].

\begin{figure}
\includegraphics[width=\linewidth]{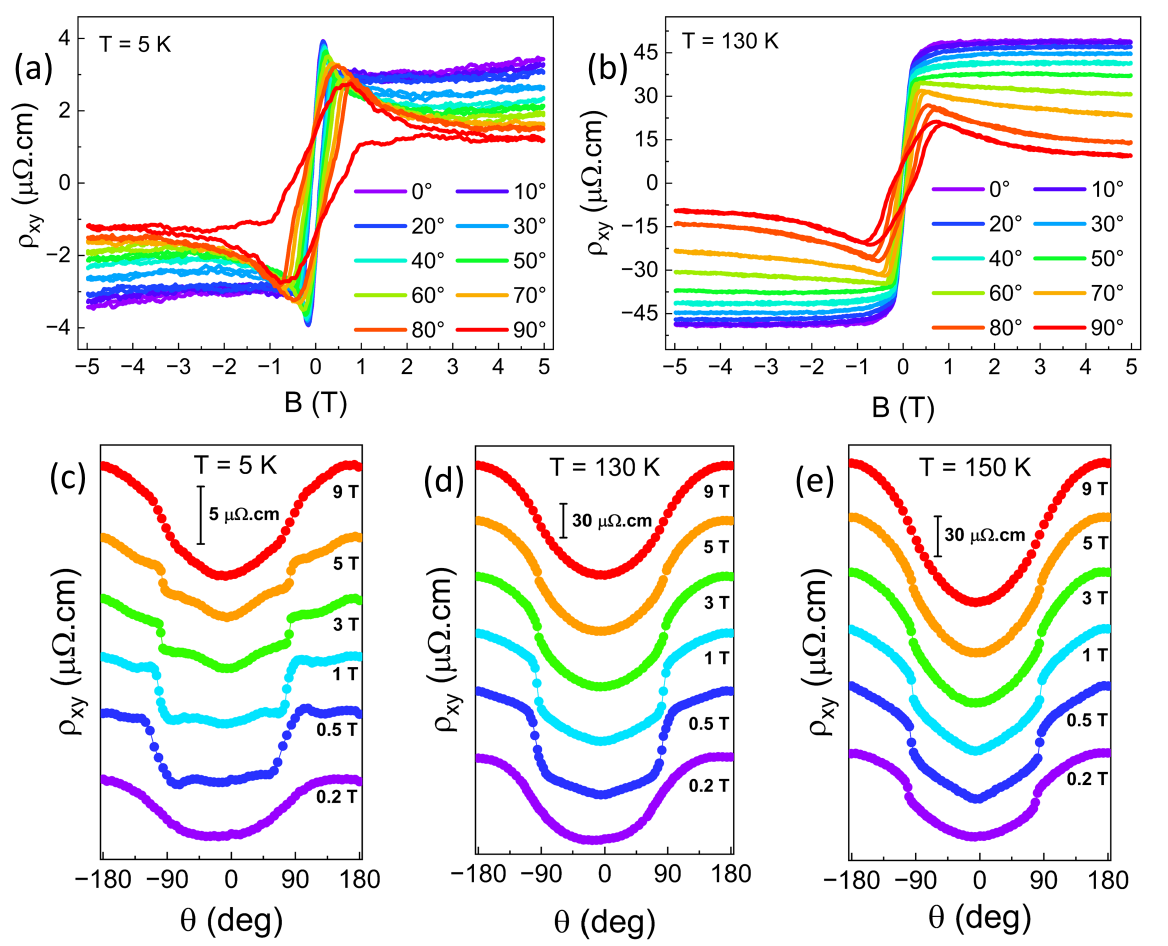}
\caption{(a) Field dependence of Hall resistivity curves at (a) 5 K and (b) 130 K for different out-of-plane orientations of the external magnetic field. Angle dependence of $\rho_{\mathrm{xy}}$ at (c) 5 K, (d) 130 K, and (e) 150 K is presented for various magnetic field strengths. At high magnetic fields (near 9 T), the curves are closer to the typical $\cos \theta$ behavior. While a low field strength, a clear deviation appears due to the presence of THE.}
\label{fig4}
\end{figure}

In general, the total Hall resistivity $\rho_{\mathrm{xy}}$ of a ferromagnetic material showing THE can be expressed as 
\begin{equation}
\rho_{\mathrm{xy}}=\rho_{\mathrm{xy}}^\mathrm{O}+\rho_{\mathrm{xy}}^\mathrm{A}+\rho_{\mathrm{xy}}^\mathrm{T}=\mathrm{R}_0 \mathrm{B}+\mathrm{R_S M}+\rho_{\mathrm{xy}}^\mathrm{T}
\end{equation}
where the ordinary, anomalous and topological Hall contribution to $\rho_{\mathrm{xy}}$ are denoted as $\rho_{\mathrm{xy}}^\mathrm{O}$, $\rho_{\mathrm{xy}}^\mathrm{A}$ and $\rho_{\mathrm{xy}}^\mathrm{T}$, respectively. $\mathrm{M}$ represents the saturation magnetization of the sample. $\mathrm{R_0}$ and $\mathrm{R_S}$ are the ordinary and anomalous Hall coefficient, respectively. We can separate the ordinary Hall component out by calculating $\mathrm{R_0}$ through a linear fit at high fields, where $\rho_{\mathrm{xy}}^\mathrm{A}$ ideally becomes constant (field-independent) beyond the saturation point and $\rho_{\mathrm{xy}}^\mathrm{T}$ becomes negligible because of coplanar and collinear spin arrangement above a certain field strength. We isolate the AHE contribution to $\rho_{\mathrm{xy}}$ by fitting a step function to model the typical AHE signal in soft ferromagnets. Fig.~\ref{fig3}(c) presents the $\rho_{\mathrm{xy}}$ curves fitted for the ordinary and anomalous contribution (see data in red line) with B$\parallel$c at 5 K and 35 K, as an illustration. The unfitted cusp-like region is the THE contribution at 5 K and is thus extracted by subtracting the fitted line from the $\rho_{\mathrm{xy}}$ curve. There is no visible topological Hall signal present in the Hall data at 35 K. Fig.~\ref{fig3}(d) presents the contour plot of the extracted THE amplitude for B$\parallel$ab at a temperature range of 2-180 K. Near $\mathrm{T_{SRT}}$, a maximum THE of approximately $\, 10 \, \mu\Omega.\mathrm{cm}$ is observed, which is significantly larger compared to most of the previous reports. The THE persists down to the lowest measured temperature of 2 K, although it rapidly vanishes above $\mathrm{T_{SRT}}$. On the other hand for B$\parallel$c, Fig.~\ref{fig3}(e) shows that the THE strength is relatively low and restricted to a narrower temperature range below 30 K. However, the THE magnitude shows little variation across this temperature range. 

Fig.~\ref{fig4}(a) and \ref{fig4}(b) present the Hall signals while rotating the magnetic field in the out-of-plane direction at 5 K and 130 K, respectively. At 5 K, a clear THE is observable in both the ab plane ($\theta=90^\circ$) and c axis ($\theta=0^\circ$) configurations as well as at the in-between polar angles. While at 130 K, the THE signal, although prominent for $\theta=90^\circ$, disappears for $\theta=0^\circ$. Additionally, the THE signal at 130 K is an order of magnitude larger compared to the same at 5 K. Fig.~\ref{fig4}(c), \ref{fig4}(d) and \ref{fig4}(e) shows the $\theta$ dependency of $\rho_{\mathrm{xy}}$ at temperatures of 5 K, 130 K, and 150 K, respectively. At high magnetic field (9 T), the angular dependence follows the typical $\mathrm{\cos\theta}$ behavior due to the presence of AHE arising from the out-of-plane component of the magnetic field $\mathrm{B\cos\theta} $. But, at the lower fields, the nature of the angular dependence deviates from this due to the contribution of the THE in the Hall signal. Notably at 5 K, this deviation is visible all the way up to 9 T, which shows the THE contribution remains significant up to a high field at low temperatures. It is worth noting from the high-field curves that the actual $\theta$ value for the c axis orientation of the magnetic field deviates marginally from $0^\circ$ due to unavoidable misalignment errors.  
The overall behavior is quite similar to the angle dependence of THE observed in the bulk Fe$_3$GeTe$_2$ (F3GT)~\cite{THE_F3GT}. 

Observation of topological spin textures like unconventional (anti)meron chains~\cite{Meron_FGT}, coexistence of (anti)skyrmions, and (anti)merons~\cite{Sk_F5GT} have been reported in the Fe$_{\mathrm{5-x}}$GeTe$_2$ with an associated THE reflected in the magneto-transport properties.
This THE originates from the deflection of charge carriers caused by an internal gauge field, which emerges from the spin chirality scalar field associated with topologically non-trivial spin textures. In addition to spin textures such as (anti)skyrmions and (anti)merons, other non-coplanar spin textures can also generate a similar gauge field due to a non-zero spin chirality. As discussed earlier, a prominent SRT in bulk as well as a multi-layer F4GT is observed which originates from the interplay of the PMA and shape anisotropy. These two anisotropies follow a different temperature dependence, giving rise to a crossover in their relative strength at $\mathrm{T_{SRT}}$. For the 78 nm thick crystal (D1), PMA dominates over the shape anisotropy below 130 K making the c axis the easy axis. This makes it easier to align the spins along the c direction leading to a collinear spin arrangement. On the other hand, it becomes harder to align the spins along the ab plane, and thus at a low field regime, a non-coplanar canted spin texture is expected to appear for the ab plane alignment of the magnetic field. Although F4GT is a centrosymmetric ferromagnet, the SRT can lead to the formation of domain structures that locally break inversion symmetry. This symmetry breaking promotes the Dzyaloshinskii-Moriya interaction (DMI)~\cite{Sk_F5GT}, leading to the emergence of noncoplanar spin textures. This results in a non-zero spin-chirality gauge field, causing a cusp in the Hall signal. At above 130 K, a large shape anisotropy takes over the PMA [Fig.~\ref{fig1}(c)] and the ab plane becomes the easy plane for most of the field orientations except along the c axis. As a result, the canted spin texture disappears, leading to absence of THE above 150 K.

 \begin{figure}
\includegraphics[width=\linewidth]{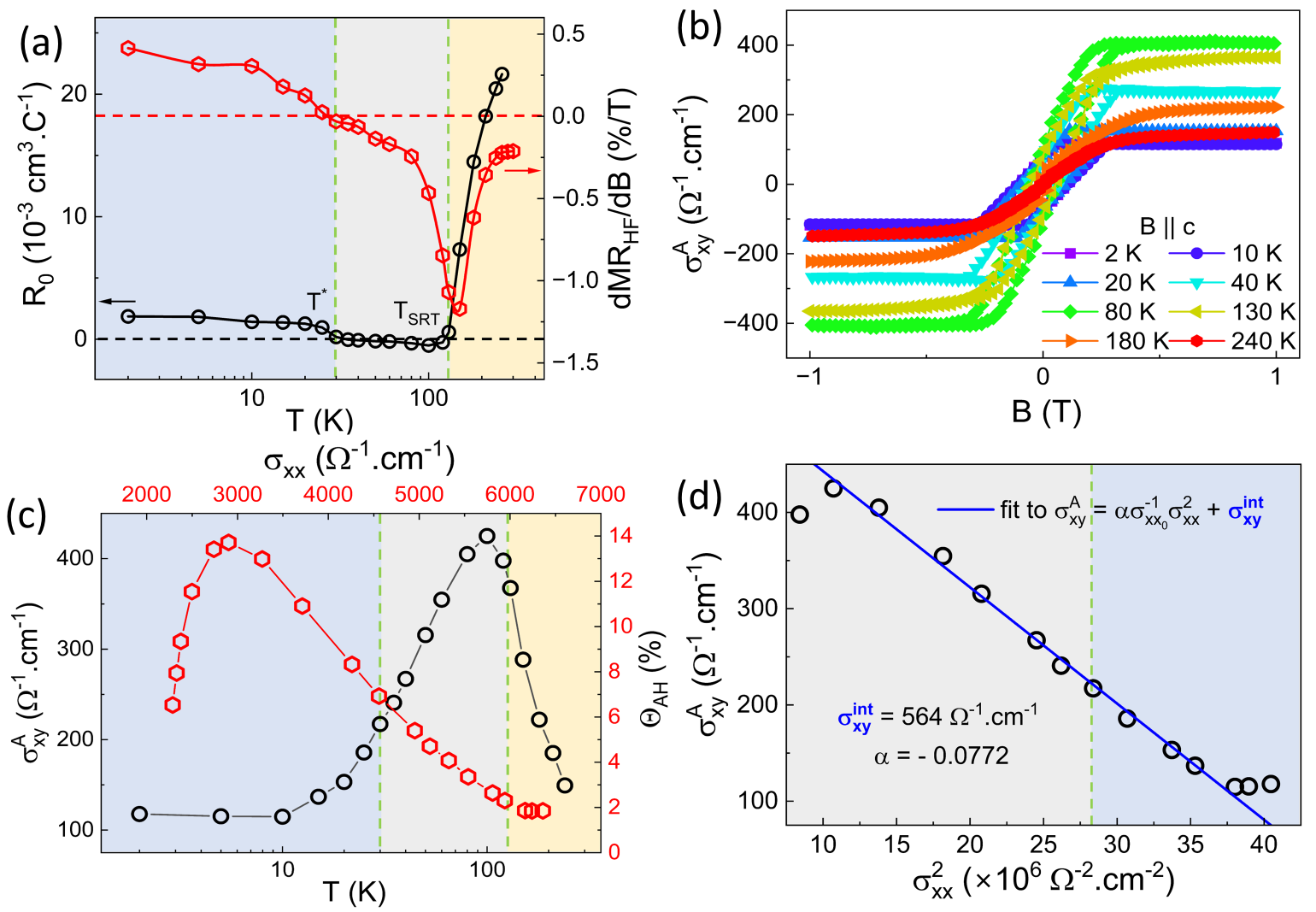}
\caption{(a) Temperature dependence of the ordinary Hall coefficient $\mathrm{R_0}$ and slope of MR vs. B curves determined from linear fitting at the high magnetic field for B$\parallel$c. From left to right, the three-color background represents the three temperature ranges: (1) below T$^*$, (2) between T$^*$ and T$_{\mathrm{SRT}}$, and (3) above T$_{\mathrm{SRT}}$, respectively. (b) Magnetic field dependence of anomalous Hall conductivity at various temperatures. (c) Temperature dependence of anomalous Hall conductivity (symbols in black) across the three temperature ranges. The red symbols present the variation of Hall angle $\Theta_{\mathrm{AH}}$ with longitudinal conductivity $\sigma_{xx}$. (d) $\mathrm{\sigma_{xy}^A}$ is plotted as a function of $\mathrm{\sigma_{xx}^2}$ for the temperature range below T$_{\mathrm{SRT}}$.} 
\label{fig5}
\end{figure}

The signature of THE observed for B$\parallel$c below 30 K is distinctive from the SRT-driven THE, both qualitatively and quantitatively. This is also unconventional in that at this low T and easy-axis orientation of the moments, the textbook scenarios that give rise to the THE are not anticipated. Field orientation being along the c axis, the frustration in spin orientation should be absent. Moreover, being far away from SRT and ferromagnet-paramagnet transition temperature, non-trivial domain formation too seems implausible. Thus, the origin of this THE most likely stems from the anomalous transition identified from the resistivity plot near 30 K (T$^*$), as discussed earlier, accompanied by a curvature change of the $M-T$ curve around the same point [Fig.~\ref{fig1}(c)]. This is also associated with a change in carrier type, as depicted in Fig.~\ref{fig5}(a) in the temperature-dependence plot of ordinary Hall coefficient ($\mathrm{R_0}$), resembling a temperature-driven reentrant Lifshitz transition. Furthermore, as we approach the low-temperature regime, the typical negative MR evolves to a positive MR (SM-Fig. 3), suggesting a strong correlation between magnetism and the Fermi surface in F4GT, similar to some other Fe-based ferromagnets~\cite{FeSn1, FeSn2}. Fig.~\ref{fig5}(a) shows the slopes of the MR curves (calculated from the high field linear regime) changing from negative to positive at around 30 K. These observations collectively point to the emergence of a complex magnetic structure below T$^*$, leading to the unconventional THE in F4GT. 

Earlier reports show thickness dependency of the domain and spin-texture formation in 2D magnets like Cr$_5$Si$_3$~\cite{Cr5Si3} and Co-doped Fe$_{5-x}$GeTe$_2$~\cite{Co_F5GT}, showing topological features in a specific range of thickness. We prepare a 2nd device (D2) with a lower thickness ($35\,\mathrm{nm}$) to look into this thickness dependency of the THE signature. SM-Fig. 4 compares the Hall signals in devices D1 and D2 at different temperatures and orientations of the external magnetic field. For ab plane orientation of the magnetic field, the SRT-driven Hall signature is visible in both cases showing a clear association with the occurrence of spin reorientation. On the other hand, for the c axis of the magnetic field, the THE signal in the low-temperature regime is absent [SM-Fig. 4(c)] in the thinner crystal (D2). This indicates that the THE signature for B$\parallel$c is sensitive to sample geometry and only appears within a specific range of crystal thickness. We observe that the reentrant Lifshitz transition in D2 is suppressed, which is likely associated with the increase in PMA strength at lower thicknesses. This enhancement in PMA strength is expected to act against the formation of non-collinear magnetic structures. Refer to Sec. V of SM~\cite{SM} for more details.

To investigate the reciprocal space properties of F4GT and the possible interconnection of real and reciprocal space topology, we calculate the field dependence of conductivity $\sigma_{\mathrm{xy}}$ using the conversion relation, $\sigma_{\mathrm{xy}}=\rho_{\mathrm{xy}}/\left(\rho_{\mathrm{xx}}^2+\rho_{\mathrm{xy}}^2\right)$. The temperature dependence of the anomalous Hall conductivity, $\sigma^{\mathrm{A}}_{\mathrm{xy}}$, obtained by subtracting the ordinary and topological Hall contributions from $\sigma_{\mathrm{xy}}$, is presented in Fig.~\ref{fig5}(b). The saturation value of $\sigma^{\mathrm{A}}_{\mathrm{xy}}$, depicted by the black symbols in Fig.~\ref{fig5}(c), reaches a maximum near T$_{\mathrm{SRT}}$, followed by a steady decrease as the temperature is lowered, eventually stabilizing at a constant value below 10 K. The anomalous Hall angle $\Theta_{\mathrm{AH}}$, defined as $\sigma^{\mathrm{A}}_{\mathrm{xy}}$/$\sigma_{\mathrm{xx}}$, attains a maximum value of nearly 14\% near T$_{\mathrm{SRT}}$ [symbols in red in Fig.~\ref{fig5}(c)]. $\sigma^{\mathrm{A}}_{\mathrm{xy}}$ plotted as a function of $\sigma^2_{\mathrm{xx}}$ shows a linear behaviour in the temperature range below T$_{\mathrm{SRT}}$. Fitting with the Tian-Ye-Jin (TYJ) scaling equation~\cite{TYJ}, 
\begin{equation}
\mathrm{\sigma_{xy}^A}=\mathrm{\alpha \sigma_{xx_0}^{-1} \sigma_{xx}^2+\sigma_{xy}^{int}}
\label{eq2}
\end{equation}
we find a large intrinsic anomalous Hall contribution, $\mathrm{\sigma_{xy}^{int}}$= 564 $\Omega^{-1}$.cm$^{-1}$. Here $\alpha$ and $\sigma_{xx_0}$ denote the skew-scattering coefficient and residual conductivity, respectively. Below 10 K, $\mathrm{\sigma_{xy}^A}$ approaches a constant value of 115 $\Omega^{-1}$.cm$^{-1}$, deviating from the quadratic $\sigma_{\mathrm{xx}}$ dependence. 

In summary, using temperature-dependent, field-angle-dependent, and thickness-dependent magneto-transport measurement we find a large topological Hall effect for the ab plane orientation of the magnetic field, which shows clear evidence of an origin associated with SRT-driven formation of non-coplanar spin-textures. This THE persists in the entire temperature range from $\mathrm{T_{SRT}}$ to the lowest measured temperature and is robust against the sample thickness variation. For the c axis orientation of the magnetic field, we find an unconventional sample size sensitive THE in the low-temperature regime, which is likely associated with a temperature-dependent reentrant Lifshitz transition. The realization of an SRT-driven large THE marks a significant advancement in our understanding of the topological properties of F4GT, in addition to establishing the same as a promising candidate for topological spintronic applications.

{\it Acknowledgements.---} SM acknowledges Department of Science and Technology, India, DST Nanomission,  for financial support. AB thanks PMRF for financial support.

\end{document}